\documentclass[lettersize,journal]{IEEEtran}
\usepackage{amsmath,amsfonts}
\usepackage{algpseudocode}
\usepackage[linesnumbered,ruled]{algorithm2e}
\usepackage{array}
\usepackage[caption=false,font=normalsize,labelfont=sf,textfont=sf]{subfig}
\usepackage{textcomp}
\usepackage{stfloats}
\usepackage{url}
\usepackage{verbatim}
\usepackage{graphicx}
\usepackage{cite}
\usepackage{multirow}
\usepackage{makecell}
\usepackage{booktabs}
\usepackage[table,xcdraw]{xcolor}
\usepackage{color}
\usepackage{arydshln}
\usepackage{bbding}
\usepackage{pifont}
\usepackage{utfsym}

\begin{document}

\title{SOLIDO: A Robust Watermarking Method for Speech Synthesis via Low-Rank Adaptation}

\author{Yue Li, Weizhi Liu, and Dongdong Lin
\thanks{Yue Li, and Weizhi Liu are with the College of Computer Science and Technology, National Huaqiao University, Xiamen 361021, China, and also with the Xiamen Key Laboratory of Data Security and Blockchain Technology, Xiamen 361021, China (e-mail: liyue\_0119@hqu.edu.cn; lwzzz@stu.hqu.edu.cn).}
}

\markboth{Journal of \LaTeX\ Class Files, July~2024}%
{Shell \MakeLowercase{\textit{et al.}}: A Sample Article Using IEEEtran.cls for IEEE Journals}


\maketitle

\begin{abstract}
The accelerated advancement of speech generative models has given rise to security issues, including model infringement and unauthorized abuse of content.
Although existing generative watermarking techniques have proposed corresponding solutions, most methods require substantial computational overhead and training costs.
In addition, some methods have limitations in robustness when handling variable-length inputs.
To tackle these challenges, we propose \textsc{SOLIDO}, a novel generative watermarking method that integrates parameter-efficient fine-tuning with speech watermarking through low-rank adaptation (LoRA) for speech diffusion models.
Concretely, the watermark encoder converts the watermark to align with the input of diffusion models. To achieve precise watermark extraction from variable-length inputs, the watermark decoder based on depthwise separable convolution is designed for watermark recovery.
To further enhance speech generation performance and watermark extraction capability, we propose a speech-driven lightweight fine-tuning strategy, which reduces computational overhead through LoRA.
Comprehensive experiments demonstrate that the proposed method ensures high-fidelity watermarked speech even at a large capacity of 2000 bps.
Furthermore, against common individual and compound speech attacks, our SOLIDO achieves a maximum average extraction accuracy of 99.20\% and 98.43\%, respectively. It surpasses other state-of-the-art methods by nearly 23\% in resisting time-stretching attacks.

\end{abstract}


\section{Introduction}
\IEEEPARstart{T}{he} prosperous growth of Artificial Intelligence-Generated Content (AIGC) has led to remarkable advancements in generative models in recent years, making text-to-speech (TTS) synthesis a highly sought-after area within this research boom.
Speech generative models, which are constructed upon Generative Adversarial Networks (GANs)~\cite{goodfellow2014gan}, transformer, and Diffusion Models (DMs)~\cite{ho2020ddpm, song2019smld, song2020ddim}, are at the forefront of advancing the naturalness of AI-generated speech, achieving unprecedented levels of humanlike vocal quality~\cite{kong2020hifigan, copet2023musicgen, kong2021diffwave, lee2022priorgrad}.
Nevertheless, these sophisticated technologies have concurrently injected certain undercurrents into what was once a placid social environment.
Malicious attackers are able to utilize open-source speech generation models to fabricate any speech they aim to disseminate, with merely the expense of training models.
Consequently, there is an urgent need for a technology capable of identifying generated content and even generative models to regulate content and protect model copyrights~\cite{CHN_POLICY, EU_POLICY, USA_POLICY}.

\begin{figure}[t]
    \centering
    \resizebox{0.98\linewidth}{!}{\includegraphics{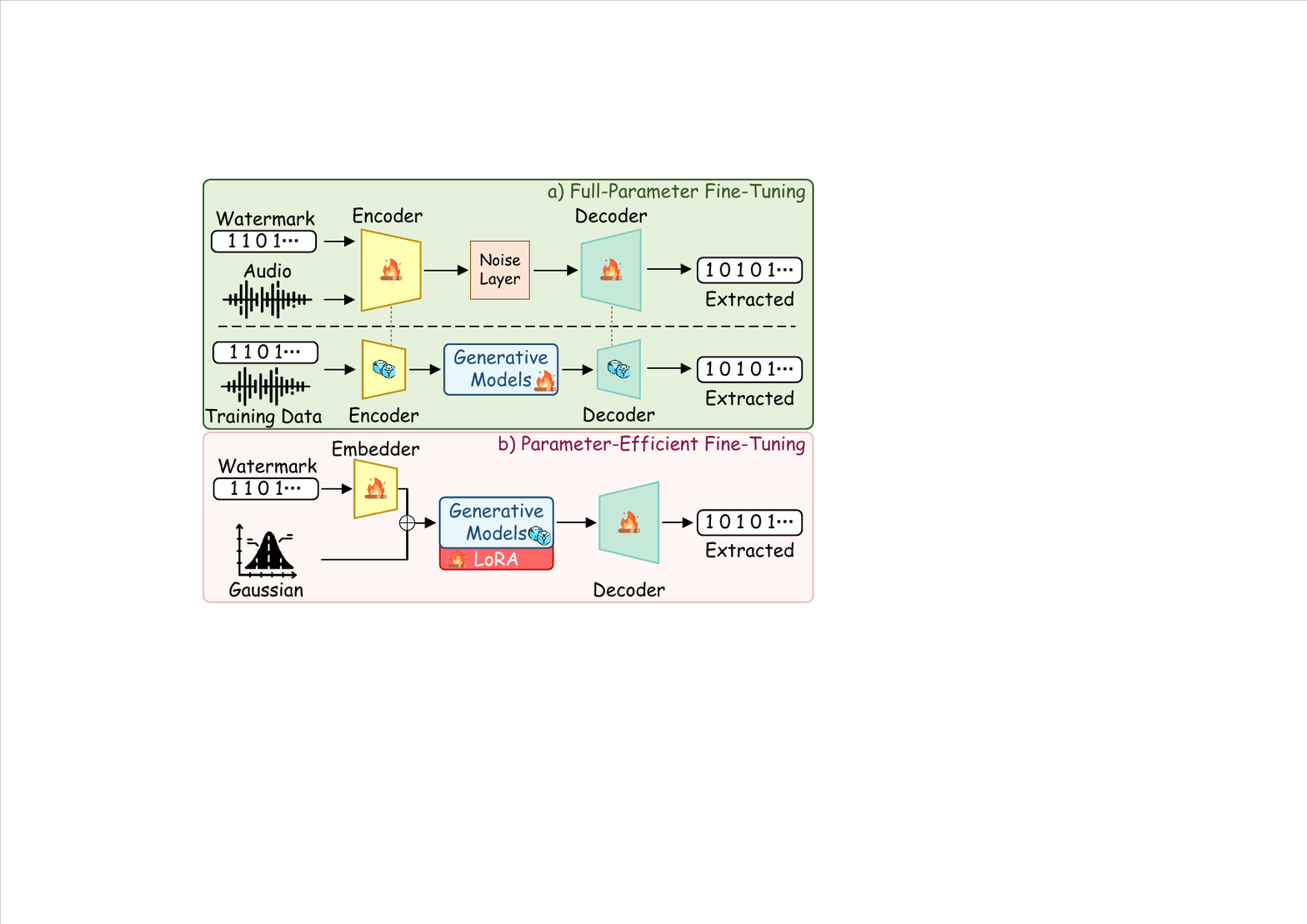}}
    \caption{Illustration of different watermarking techniques with full-parameters and parameter-efficient fine-tuning strategy. The proposed method leverages low-rank adaptation to enable single-step model training while drastically reducing the number of trainable parameters.}
    \label{fig_overview}
    \vspace{-0.3cm}
\end{figure}

Watermarking technology, due to its traceability, is often used as a proactive measure for copyright protection~\cite{ren2024sok, zhao2024sok, li2021survey}.
Previous studies primarily utilized post-hoc watermarking techniques for protecting speech intellectual property, which can be categorized into handcrafted-based and deep learning-based watermarking methods.
Handcrafted-based watermarking typically designs specific algorithms tailored to speech features for watermark embedding~\cite{saadi2019normspace, zhao2021desyn-fsvcm, natgunanathan2017patchwork, liu2018patchwork1}, whereas deep learning-based watermarking integrates the watermark with speech representations through neural networks~\cite{roman2024audioseal, chen2023wavmark, liu2024timebre, liu2023dear}.
Although post-hoc watermarking can effectively trace natural or generated content, it falls short in tracing the generative models. 
Consequently, this has spurred the development of generative watermarking capable of safeguarding the copyright of models~\cite{cho2022attributable, juvela2024collaborative, zhou2024traceablespeech, cheng2024hifiwm, san2025latent, liu2024groot}, which can be categorized into \textit{parameter training-based watermarking} and \textit{parameter frozen-based watermarking} according to their training strategy.
In the first category of generative watermarking, existing approaches root the watermark into the generative model through \textit{full-parameter training (FPT)}, including designing specific loss functions for training~\cite{cho2022attributable, juvela2024collaborative} and employing a two-stage training strategy~\cite{zhou2024traceablespeech, cheng2024hifiwm}.
The second category of generative watermarking employs the parameter-frozen strategy (PT) for the generative model, where only the watermark encoder and decoder is trained to achieve watermarking~\cite{liu2024groot}.

However, the aforementioned methodologies confront persistent limitations that require resolutions.
On the one hand, in terms of \textit{computational overhead}, existing parameter training-based watermarking methods that train models through FPT incur significant computational resource consumption, and the two-stage training strategy further increases training costs and time.
In the field of image watermarking, preliminary research has already explored instilling the watermark into generative models utilizing \textit{additional parameter training (APT)}, including approaches such as utilizing LoRA~\cite{hu2021lora} for watermarking~\cite{fengaqualora, lin2024ewlora}.
Nevertheless, such APT strategies have yet to be applied to speech watermarking, leaving the field still grappling with the challenge of high computational resource consumption.
On the other hand, in terms of \textit{robustness}, although existing watermarking methods maintain competent extraction performance against common speech attacks, certain deep learning-based approaches (whether post-hoc or generative) lack the inherent capability to process variable-length inputs directly. This limitation compromises their effectiveness when confronting variable-length attacks such as rear-segment cropping and time-stretching attacks.

To eliminate the limitations, we propose a generative speech watermarking method via LoRA.
This method is the first to incorporate parameter-efficient fine-tuning (PEFT) into speech watermarking, effectively reducing computational overhead while establishing a traceable mechanism for both models and content.
As illustrated in Fig.~\ref{fig_overview}, the proposed method requires training only the watermark encoder-decoder and additional parameters (\textit{e.g.}, LoRA) to achieve watermark embedding.
Our SOLIDO accomplishes the training of generative models with significantly fewer parameters than FPT and eliminates the cumbersome two-stage training process.
To mitigate the impact of watermark embedding on model performance, the watermark encoder, comprising only three operations, is designed to transform watermarks into latent variables aligned with the diffusion model's input.
Moreover, we construct the watermark decoder employing depthwise separable convolutions, enabling it to better capture fine-grained features from speech waveforms for high-precision watermark recovery. 
Notably, this decoder can accept variable-length inputs, allowing watermark extraction even from arbitrary-length speech without extra processing.
In addition, to further boost watermarking performance, we propose a speech-driven lightweight fine-tuning strategy, which achieves high-fidelity watermarked audio generation at a lower computational cost.

The main contributions are summarized as follows:
\begin{itemize}
    \item We innovatively incorporate parameter-efficient fine-tuning with speech watermarking and propose a generative watermarking via low-rank adaptation, which enables both copyright protection for diffusion models and authentication of generated content.

    \item To preserve the original model's performance and watermarking effectiveness, the lightweight watermark encoder and decoder are meticulously designed. Notably, the decoder is capable of accurately recovering watermarks directly from variable-length inputs.

    \item To balance speech quality and watermark extraction accuracy, we further propose a speech-driven lightweight fine-tuning strategy (SDFT), where the computational overhead of watermark training is reduced through LoRA.

    \item Extensive experiments demonstrate that the proposed method achieves high-fidelity speech generation with a large capacity of 2000 bps. In addition, comparative experiments highlight its robustness against various speech attacks, \textit{e.g.}, common attacks and variable-length attacks.

\end{itemize}

\section{Related Work}
\subsection{Post-hoc Watermarking}
\label{sec_rw_phw}
Post-hoc watermarking methods embed identifying marks into natural or AI-generated speech for copyright protection. These methods can be divided into two main categories: handcrafted-based watermarking and deep learning-based watermarking approaches.

\textbf{Handcrafted-based watermarking methods} Handcrafted watermarking methods primarily focus on designing corresponding algorithms for watermark embedding and extracting based on speech features.
Since the frequency-domain features of speech exhibit better robustness compared to time-domain features, they are often utilized as the embedding features for watermarking~\cite{natgunanathan2017patchwork, saadi2019normspace, zhao2021desyn-fsvcm}.
Using the Discrete Cosine Transform (DCT), Natgunanathan et al.~\cite{natgunanathan2017patchwork} adopted a patchwork-based approach, embedding the watermark into the DCT coefficients of multiple layers of speech.
Using Arnold transformation, Saadi et al.~\cite{saadi2019normspace} first encrypted the watermark. Subsequently, the watermark is embedded into the norm space of speech after applying discrete wavelet transform and DCT.
Zhao et al.~\cite{zhao2021desyn-fsvcm} utilized Singular Value Decomposition to extract frequency singular value coefficients from the DCT coefficients of speech and then embedded the watermark into them.

\textbf{Deep learning-based watermarking methods} achieve the integration of watermarks and audio features through neural networks, enhancing watermark performance by designing different network architectures~\cite{chen2023wavmark, roman2024audioseal, liu2024timebre}.
For time-domain features of speech, Roman et al.~\cite{roman2024audioseal} obtain discrete features using the speech neural codec and fuse them with the watermark to reconstruct watermarked audio.
In addition, leveraging the invertible neural network, Chen et al.~\cite{chen2023wavmark} complete watermark embedding by fusing the short-time Fourier transform (STFT) features of the watermark and speech.
To defend against voice cloning attacks, Liu et al.~\cite{liu2024timebre} repeat the watermark and combine it with the STFT magnitude features of the speech, then apply the inverse STFT to obtain watermarked audio.

\subsection{Generative Watermarking}
\label{sec_gw}
Unlike post-hoc watermarking techniques that modify content after generation, generative watermarking represents a paradigm shift by integrating the watermarking into the generating process of generative models.
Contemporary watermarking techniques can be classified into two categories: parameter training-based~\cite{cho2022attributable, juvela2024collaborative, zhou2024traceablespeech, cheng2024hifiwm} and parameter frozen-based watermarking methods~\cite{liu2024groot}.

\textbf{Parameter training-based watermarking} indicates that watermark embedding is achieved by performing FPT. 
Cho et al.~\cite{cho2022attributable} retrained the generative model with specialized constraints, ensuring that the retrained model could trace its synthesized speech.
Juvela et al.~\cite{juvela2024collaborative} jointly trained the generative model and the detector, enabling the classifier to identify speech synthesized by this specific model.
However, the aforementioned methods are zero-bit watermarking schemes, which can only determine the presence of a watermark based on content rather than extracting specific embedded information.
Therefore, subsequent research has shifted focus toward multi-bit watermarking.
Zhou et al.~\cite{zhou2024traceablespeech} and Cheng et al.~\cite{cheng2024hifiwm} utilized a two-stage training strategy for watermark embedding. Specifically, the watermark encoder-decoder is first pretrained, and then the generative model is fine-tuned using the pretrained encoder and decoder.

\textbf{Parameter frozen-based watermarking} means that the generative model does not need to participate in training. Following this principle, Liu et al.~\cite{liu2024groot} utilized JOPT to integrate the watermarking process with the generation process while keeping the generative model parameters frozen.

In a nutshell, post-hoc speech watermarking, due to its decoupling from generative models, cannot trace the models themselves and focuses solely on content. 
Most existing generative watermarking methods employ FPT strategy for watermarking, resulting in substantial computational overhead. 
Besides, in the speech domain, research on watermarking using APT has not yet been explored.
Therefore, the proposed method integrates PEFT techniques and leverages APT to reduce computational overhead and training costs, making it more aligned with practical application requirements.

\section{Preliminaries}
\label{sec_pre}
\textbf{Diffusion Denoising Probabilistic Model (DDPM)}~\cite{ho2020ddpm}. The proposed \textsc{solido} focuses on DDPM-based vocoders, specifically DiffWave~\cite{kong2021diffwave} and PriorGrad~\cite{lee2022priorgrad}, to generate watermarked speech.
A brief overview of DDPMs in the context of speech generation is provided below.

In the diffusion process of DDPM, the original input $\mathbf{s}_t$ of the DDPM is obtained by adding Gaussian noise to the original speech $\mathbf{s}_0 \sim q_{data}(\mathbf{s}_0)$ step by step, with the standard deviation of noise determined by hyper-parameter $\beta_t \in (0,1)$:
\begin{gather}
    q(\mathbf{s}_{t}|\mathbf{s}_{t-1}) = \mathcal{N}(\mathbf{s}_{t}; \sqrt{1-\beta_{t}}\, \mathbf{s}_{t-1}, \beta_{t}\mathbf{I}), \\[1ex]
    q(\mathbf{s}_{1:T}|\mathbf{s}_{0}) = \prod_{t=1}^{T} q\left(\mathbf{s}_{t}|\mathbf{s}_{t-1}\right),
\end{gather}
where $\mathbf{I}$ means identity matrix.
Furthermore, let $\alpha_t = 1 - \beta_t$, $\overline \alpha_t = \prod_{i=1}^t \alpha_i$ and $\mathbf{\epsilon} \sim \mathcal N(\mathbf{0}, \mathbf I)$, then $\mathbf{s}_t$ can be acquired by simply adding noise in a single step: 
\begin{equation}
    \mathbf{s}_t = \sqrt{\overline{\alpha}_t} \mathbf{s}_0 + \sqrt{1-\overline{\alpha}_t} \mathbf{\epsilon},
\end{equation}
 
The denoising process involves using a neural network to approximate the noise added during the diffusion process, which can be expressed as:
\begin{gather}
    p_\theta(\mathbf{s}_{0:T}) = p(\mathbf{s}_T) \prod_{t=1}^{T} p_\theta(\mathbf{s}_{t-1}|\mathbf{s}_t), \\[1ex]
    p_\theta(\mathbf{s}_{t-1}|\mathbf{s}_t) = \mathcal{N}(\mathbf{s}_{t-1}; \mu_\theta(\mathbf{s}_t, t), \, \Sigma_\theta(\mathbf{s}_t, t)),
\end{gather}
where $\theta$ denotes the learnable parameter. Concretely, this process aims to remove the noise from latent variable $\mathbf{s}_t$ by employing the prediction network $\mathbf{\epsilon}_\theta$ to estimate the noise added during the diffusion process step by step. The completed process can be calculated as:
\begin{equation}
    \mathbf s_{t-1} = \frac{1}{\sqrt{\alpha_t}} \bigg(\mathbf s_t - \frac{1-\alpha_t}{\sqrt{1-\overline \alpha_t}} \mathbf{\epsilon}_\theta(\mathbf s_t, t, c) \bigg) + \mathbf{\delta}_t \mathbf z,
\end{equation}
where $\delta_t\mathbf{z}$ denotes the random noise, $\mathbf{z} \sim\mathcal N(\mathbf{0}, \mathbf{I})$, and $c$ is the mel-spectrogram.

The training of $\mathbf{\epsilon}_\theta$ aims to fit the noise $\epsilon$. Thus, the parameters $\theta$ need to be continuously learned and updated by maximizing the variational lower bound. Therefore, the final objective function can be simplified as:
\begin{equation}
\label{eq_train}
    \mathcal L_{simple} = \mathbb E_{\mathbf{s}_0, t, \epsilon} \big[ ||\epsilon - \epsilon_\theta(\mathbf{s}_t, t, c) ||^2 \big].
\end{equation}

\textbf{Low-Rank Adaptation.} 
LoRA~\cite{hu2021lora}, as one of the PEFT techniques, its core idea is to freeze the original parameters of the model and introduce two additional low-rank matrices for training. This low-rank decomposition of matrices can be viewed as a form of approximate numerical decomposition technique.

Given the parameters $W_G$ of the generative model, the number of parameters required for conventional full fine-tuning is $W_G$.
When using LoRA, an additional parameter $\Delta W$ is introduced on top of $W_G$. 
During subsequent training, the original parameters $W_G$ remain frozen, and only the trainable parameters $\Delta W$ are updated.
Since the core of LoRA lies in the use of low-rank decomposition, the trainable parameters $\Delta W$ can be further decomposed into two low-rank matrices, $A$ and $B$, where $A \in \mathbb{R}^{d \times r}$, $B \in \mathbb{R}^{r \times k}$, and the rank $r \ll \min(d, k)$.
Therefore, the updated formula for the total model parameters $W$ after fine-tuning using LoRA can be expressed as:
\begin{equation}
    W = W_G + \Delta W = W_G + \alpha BA.
\end{equation}
where $\alpha$ is a scale factor. Setting $\alpha$ helps reduce the need for readjusting hyperparameters when $r$ changes.

When initializing the two low-rank matrices, matrix $A$ is typically initialized using Gaussian initialization, while matrix $B$ is initialized as a zero matrix. 
This initialization strategy ensures that the training starts with $\Delta W = 0$, thereby preventing any impact on the model's original output.
In summary, when using LoRA, the training process of the generative model only computes gradients for the two low-rank matrices $A$ and $B$, while freezing the majority of the original parameters $W_G$, significantly reducing training time and resource consumption. 
In addition, during training, $\alpha$ is set for $W$ to help eliminate the impact of changes in $r$ on parameter updates, ensuring more consistent hyperparameter tuning.



\section{The Proposed Method}

\begin{figure*}
    \centering
    \resizebox{0.9\textwidth}{!}{\includegraphics{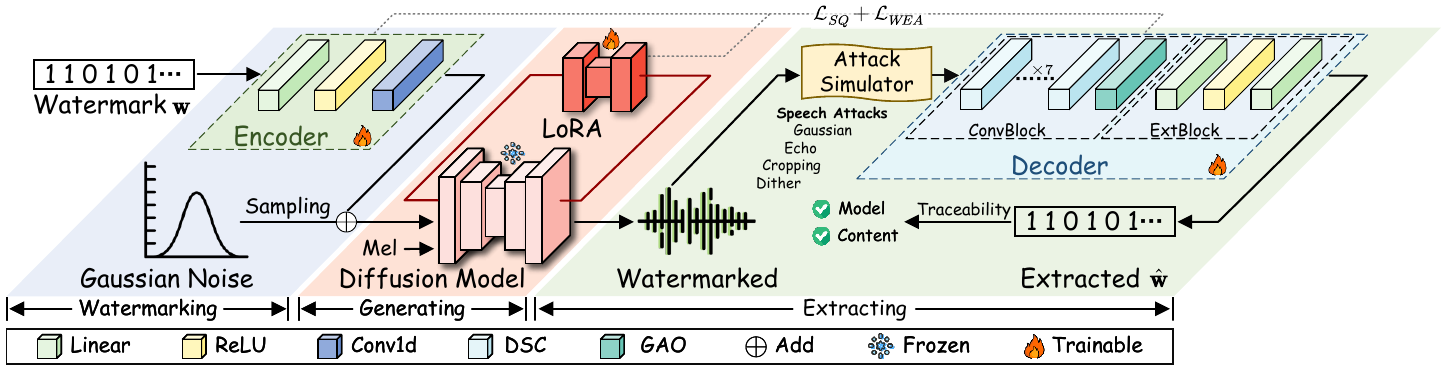}}
    \caption{\textbf{The pipeline of the proposed \textsc{solido}.} In the watermarking phase, the watermark encoder encodes the watermark into the latent variable, which is then combined with the original Gaussian-sampled input to acquire modified input. The diffusion model with LoRA takes this modified input to generate the watermarked speech in the generating phase. During the watermark extracting phase, the watermarked speech initially undergoes the attack simulator and is subsequently fed into the watermark decoder to recover the watermark.}
    \label{fig_pipe}
\end{figure*}

\subsection{Overview}
\label{sec_overview}
The proposed method aims to address three fundamental challenges in the field: 
(1) establishing robust traceability for the diffusion model and its generative content, 
(2) developing flexible processing capabilities for variable-length inputs, 
and (3) reducing the steps of algorithm construction and computational overhead of watermark training.
Therefore, the proposed \textsc{soildo}, as depicted in Fig.~\ref{fig_pipe}, comprises three phases designed to tackle these issues.
First, the watermark encoder $\mathbf{Enc}(\cdot)$ is employed to transform the watermark $\mathbf{w}$ into the latent variable $\sigma$. 
Then, the original input of model $\mathbf{s}_T$ is combined with $\sigma$ and fed into the diffusion model to generate the watermarked speech $\mathbf{\hat{s}_0}$. 
Subsequently, $\mathbf{\hat{s}_0}$ undergoes the attack simulator $\mathbf{AS}(\cdot)$. 
Finally, the watermark decoder $\mathbf{Dec}(\cdot)$ recovers the watermark $\hat{\mathbf{w}}$ from the watermarked speech.
It is emphasized that the parameters of the diffusion model are frozen during training, and the model is updated solely through LoRA.
The details will be provided in the following sections.

\subsection{The Pipeline of the Proposed Method}
\label{sec_wm}

\begin{figure}
    \centering
    \includegraphics[width=\linewidth]{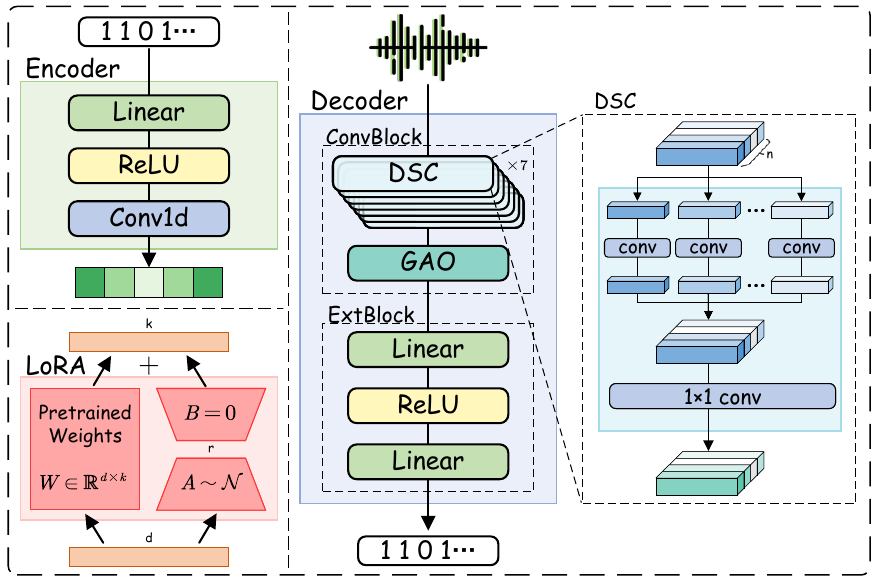}
    \caption{Architecture of SOLIDO.}
    \label{fig_archi}
\end{figure}

\subsubsection{Watermarking Phase}
\label{sec_watermarking}
During this phase, the primary objective is to inject the watermark into the diffusion model while preserving the model's intrinsic performance.
To realize this, the watermark encoder is designed to convert the watermark into the latent variable, which is subsequently combined into the original input of the diffusion model.
This procedure also necessitates that the latent variable be adjusted to approximate the original input as closely as possible, ensuring that the addition operation does not disrupt the normal operation of the diffusion model.
To achieve the model lightweighting, the encoder's architecture is deliberately streamlined, consisting of three essential components: a linear layer, a ReLU activation, and a one-dimensional convolution layer (Conv1d), as depicted in Fig.~\ref{fig_archi}. 
The linear layer maps the watermark of arbitrary length to a hidden representation that corresponds dimensionally with the diffusion model's original input.
Subsequently, the hidden representation undergoes normalization through ReLU.
Finally, the Conv1d layer captures the fine-grained features of the watermark, effectively minimizing the performance impact of the watermarking process.

Specifically, given the watermark $\mathbf{w} \in \{0, 1\}^l$, where $l$ denotes the length of the watermark, the watermark encoder $\mathbf{Enc}(\cdot)$ first transforms $\mathbf{w}$ into the latent variable $\sigma$, which is then combined with the original input $\mathbf{s}_T$ of the diffusion model to acquire the modified input $\hat{\mathbf{s}}_T$. The complete watermarking phase can be expressed as:
\begin{gather}
\label{eq_watermarking}
\sigma = \mathbf{Enc}(\mathbf{w}), \\[1.5ex]
\hat{\mathbf{s}}_T = \sigma + \mathbf{s}_T \in \mathbb{R}^{B \times C \times L},
\end{gather}
where $B$ is batch size, $C$ denotes the channel of speech and $L$ represents the length of the original input.

\subsubsection{Generating Phase}
\label{sec_generating}
The generation phase utilizes the diffusion model to synthesize watermarked speech based on modified input $\hat{\mathbf{s}}_T$.
Although the preceding watermarking phase has been optimized to minimize performance degradation, this operation inevitably introduces additional noise to the diffusion model.
In light of this, this phase implements LoRA technology, which maintains the original parameters of the diffusion model in a frozen state while adapting to the watermarking operation through the introduction of a minimal set of supplementary parameters.

In the training process of this method, the total parameters $W$ of the diffusion model are divided into two parts: the original parameters $W_G$ and the trainable parameters $\Delta W$, denoted as $W = W_G + \Delta W$.
Therefore, during the generation phase, the diffusion model first takes $\hat{\mathbf{s}}_T$ as input and the mel-spectrogram $c$ of the speech as the condition, synthesizing the watermarked speech through the inclusion of these additional trainable parameters. 
The complete generating phase is:
\begin{equation}
\label{eq_generating}
\hat{\mathbf{s}}_0 = \mathcal{G}(\hat{\mathbf{s}}_T,t,c; W_G + \Delta W),
\end{equation}
where $\mathcal{G}(\cdot)$ represents the diffusion model and $t$ denotes the diffusion step. 
After integrating LoRA into the diffusion model, only the trainable parameters $\Delta W$ require gradient optimization, while the original parameters $W_G$ of the model remain frozen.
Furthermore, to balance the watermarked speech quality and watermark extraction accuracy, this method employs a speech-driven lightweight fine-tuning strategy for training, which will be detailed in Section~\ref{sec_sdft}.
In addition, during the inference process, the diffusion model utilizing LoRA eliminates the need for extra computational overhead when performing generation tasks. 

\begin{table*}[]
\centering
\caption{Specific settings and descriptions for each type of attack, including the parameters and \\ probabilities used in the attack simulator and the parameters utilized for robustness experiment during evaluation.}
\setlength{\tabcolsep}{1mm}
\resizebox{0.92\textwidth}{!}{
\begin{tabular}{ccccl}
\toprule
Attack Type               & Param.         & Prob. & Param. (Infer) & Description                                                                         \\ 
\midrule
\multicolumn{5}{l}{\textit{Non-Attack}}                                                                                                                   \\
Non                       & -              & 0.2   & -              & Watermarked speech without attacking.                                                \\ 
\midrule
\multicolumn{5}{l}{\textit{Speech Post-process Attacks}}                                                                                                  \\
Gaussian Noise (GN)       & 15-20 dB       & 0.2   & 5/10/15/20 dB  & Adding white Gaussian noise to the watermarked speech.                               \\
Echo                      & default        & 0.2   & default        & Attenuating the watermarked speech volume by a factor of 0.4, delaying it by 100ms.  \\
Rear-Segment Cropping (RSC)                 & 25\%  & 0.2   & 50\% & Cropping initial and final of the watermarked speech based on cut-rate.              \\
Dither                    & RPDF           & 0.2   & TPDF           & Applying eliminating nonlinear truncation distortion to the watermarked speech.      \\
Low-Pass Filtering (LPF)  & -              & -     & 3 kHz          & Permitting watermark speech signals below the threshold to pass.                      \\
Band-Pass Filtering (BPF) & -              & -     & 0.3-8 kHz      & Allowing watermark speech signals within the minimum and maximum thresholds to pass. \\
Pink Noise (PN)           & -              & -     & 0.5            & Adding pink background noise to the watermarked speech.                              \\ 
\bottomrule
\end{tabular}
}
\label{tab_attack}
\end{table*}

\subsubsection{Attack Simulator}
\label{sec_as}
To simulate potential attacks that could be confronted in real-world scenarios, we introduced an attack simulator prior to the watermark extracting stage during the training phase, which aims to boost the robustness of the proposed method. 
Given a watermarked speech $\hat{\mathbf{s}}_0$, it initially passes through the attack simulator $\mathbf{AS}(\cdot)$ to yield an attacked speech $\mathbf{AS}(\hat{\mathbf{s}}_0)$ and is then input into the watermark decoder.
Considering the impracticality of simulating all potential attacks encountered in real-world environments, we incorporated three prevalent attack types (\textit{Gaussian Noise}, \textit{echo}, and \textit{dither}) complemented by a \textit{rear-segment cropping} attack into our attack simulator as representative speech post-processing operations, as illustrated in Table~\ref{tab_attack}.
To mitigate overfitting to any particular attack, we assigned equal probabilities of 0.2 to each of the four attacks. In addition, we incorporated a \textit{Non-attack} operation with an equivalent probability of 0.2, whereby the speech signal remains unmodified. This uniform probability distribution ensures balanced exposure to all five operations during training.
Notably, we initially configured the parameters of these attacks at low intensities. However, these attacks are implemented with more aggressive parameter settings for validating in robustness experiments.

\subsubsection{Extracting Phase}
\label{sec_extracting}
To establish reliable traceability of the diffusion model and synthesized speech while accommodating variable-length inputs, a watermark decoder is designed meticulously to facilitate efficient watermark recovery during the extracting phase.
The watermark decoder architecture integrates a convolutional block (ConvBlock) and an extraction block (ExtBlock), as depicted in Fig.~\ref{fig_archi}.
We construct the ConvBlock utilizing seven depthwise separable convolution (DSC) layers combined with a global averaging operation (GAO). 

The ExtBlock is composed of one linear layer, a ReLU activation function, and another linear layer. The final linear layer outputs the extracted watermark.
In contrast to the standard convolutional layer, DSC applies an individual filter to each input channel, decomposing the input feature into distinct separate layers, one for filtering and another for combining~\cite{chollet2017xception, howard2017mobilenets, sandler2018mobilenetv2}.
As a single-channel signal, speech leverages DSC's channel-wise feature extraction capabilities to capture enhanced temporal representations. 
After the ConvBlock completes feature extraction, the hidden features undergo a GAO.
This operation computes the mean across the feature dimensions, aggregating hidden features of arbitrary speech lengths into a unified scalar value.
In this way, a fixed-length feature is obtained regardless of variations in the input length.
It is also incorporated because it eliminates the dependency of certain watermarking methods on input dimensions when using convolutional layers, enabling the watermark decoder to handle variable-length inputs.
Besides, the averaging process preserves global features, ensuring the integrity of the watermark features.
Subsequently, the ExtBlock recovers the watermark from the averaged hidden feature.

upon receiving the watermarked speech $\hat{\mathbf{s}}_0$, it is first fed into the attack simulator $\mathbf{AS}(\cdot)$ to generate the attacked speech $\mathbf{AS}(\hat{\mathbf{s}}_0)$.
The attacked speech is then processed by the watermark decoder $\mathbf{Dec}(\cdot)$, where it first passes through the ConvBlock $\mathbf{CB}(\cdot)$ to extract hidden features $h$, which can be expressed as:
\begin{equation}
    h = \mathbf{CB}(\mathbf{AS(\hat{\mathbf{s}}_0)}).
\end{equation}
Then, $h$ is aggregated via GAO ($\mathbf{GAO}(\cdot)$) to obtain the fixed-length feature $h_{fix}$, enabling the effective accommodating of variable-length inputs:
\begin{equation}
    h_{fix} = \mathbf{GAO}(h) \in \mathbb{R}^{B \times l_{fix}},
\end{equation}
where $l_{fix}$ denotes the fixed length obtained after aggregating inputs of arbitrary lengths.
Finally, $h_{fix}$ is fed into the ExtBlock $\mathbf{Ext}(\cdot)$ for watermark recovery, which can be represented as:
\begin{equation}
    \hat{\mathbf{w}} = \mathbf{Ext}(h_{fix}) \in \mathbb{R}^{B \times l}.
\end{equation}


\subsection{Speech-Driven Lightweight Fine-tuning Strategy}
\label{sec_sdft}
The primary purpose of our \textsc{solido} is to reduce the computational overhead associated with model training while simultaneously balancing watermarked speech quality and watermark extraction accuracy. To meet this goal, we further proposed a speech-driven lightweight fine-tuning strategy (SDFT).
This strategy implements LoRA to fine-tune the diffusion model, introducing only a small number of additional parameters while keeping the original parameters of the diffusion model frozen.
Unlike the training objective function of DDPM, this strategy drives gradient updates in the model by constraining the distance between the original speech and the watermarked speech.

As described in Sec.~\ref{sec_generating}, the overall parameters $W$ of the diffusion model are composed of the original parameters $W_G$ and trainable parameters $\Delta W$. 
Specifically, $\Delta W$ can be decomposed into two low-rank matrices, $A$ and $B$, where $\Delta W = BA \in \mathbb{R}^{d \times k}$, $A \in \mathbb{R}^{d \times r}$, and $B \in \mathbb{R}^{r \times k}$, as described in Sec.~\ref{sec_pre}.
Therefore, according to the generation process outlined in Eq.~\ref{eq_generating}, the specific denoising process can be refined as follows:
\begin{equation}
    \hat{\mathbf{s}}_{t-1} = \frac{1}{\sqrt{\alpha_t}} \left( \hat{\mathbf{s}}_t - \frac{1 - \alpha_t}{\sqrt{1 - \alpha_t}} \epsilon_W \left( \hat{\mathbf{s}}_t, t, c \right) \right) + \delta_t \mathbf{z}.
\end{equation}

Once the watermarked speech is generated by the diffusion model augmented with LoRA, the diffusion model can then be trained by employing the corresponding loss function.
To enhance the quality of the watermarked speech, the mel-spectrogram loss $\mathcal{L}_{M}$ is first utilized to constrain the distance between the original speech and the watermarked speech, which can be expressed as:
\begin{equation}
    \mathcal{L}_{M} = || \phi(\mathbf{s}_0) - \phi(\hat{\mathbf{s}}_0) ||_1,
\end{equation}
where $\mathbf{s}_0$ is original speech, $\hat{\mathbf{s}}_0$ represents the watermarked speech, $\phi(\cdot)$ denotes the function of mel-sepctrogram transformation, and $|| \cdot ||_1$ means the $L_1$ norm.
In addition, we utilize the Short-Time Fourier Transform (STFT) magnitude loss $\mathcal{L}_{S}$ to further boost the quality of the watermarked speech. It can be represented as:
\begin{equation}
    \mathcal{L}_{S} = || \log(\xi(\mathbf{s}_0)) - \log(\xi(\hat{\mathbf{s}}_0)) ||_1,
\end{equation}
where $\xi(\cdot)$ denotes the function of STFT magnitude transformation. Therefore, the overall loss for promoting the watermarked speech quality can be defined as:
\begin{equation}
    \mathcal{L}_{SQ} = \lambda_{m} \mathcal{L}_{M} + \lambda_{s} \mathcal{L}_{S},
\end{equation}
where $\lambda_{m}$ and $\lambda_{s}$ are hyper-parameters of $\mathcal{L}_{M}$ and $\mathcal{L}_{S}$, used to maintain the balance between the two terms.

On the other hand, binary cross-entropy is employed to elevate the accuracy of watermark extraction, and this loss $\mathcal{L}_{WEA}$ can be expressed as:
\begin{equation}
    \mathcal{L}_{WEA} = - \sum_{i=1}^{k} w_i \log {\hat{w}_i} + (1-w_i) \log(1-\hat{w}_i).
\end{equation}
In a nutshell, the final objective function is represented as:
\begin{equation}
    \mathcal{L} = \mathcal{L}_{SQ} + \lambda_{wea} \mathcal{L}_{WEA},
\end{equation}
where $\lambda_{wea}$ is a hyper-parameter to strike a balance between the speech quality and extraction accuracy.
The complete process of speech-driven lightweight fine-tuning strategy is presented in Algorithm~\ref{algo}.

\newcommand{\lIfElse}[3]{\lIf{#1}{#2 \textbf{else}~#3}}
\begin{algorithm}[t]
\caption{Speech-Driven Lightweight Adaption Fine-tuning Strategy (SDFT).}
\label{algo}
\SetKwInOut{Require}{Require}
\SetKwInOut{Output}{Output}

\Require{Watermark $\mathbf{w}$, watermark encoder $\mathbf{Enc}(\cdot)$ and decoder $\mathbf{Dec}(\cdot)$ with their trainable parameters $\theta_{enc}$ and $\theta_{dec}$, diffusion model $\epsilon_{\theta}$ with trainable parameters $\Delta W$ and frozen weights $W_G$ of DM, hyper-parameter $\alpha_t$, diffusion step $T$, attack simulator $\mathbf{AS}(\cdot)$, and objective function $\mathcal{L}_{SQ}$ and $\mathcal{L}_{WEA}$.}

\Repeat{converged}{

$\mathbf{s}_T \leftarrow \mathbf{s}_T \sim \mathcal{N}(0, \mathbf{I})$;

\tcp{Watermarking}
$\sigma = \mathbf{Enc}(\mathbf{w})$; \Comment $\mathbf{w} = \{ (w_i), w_i \in \{0, 1\}\}_{i=1}^l $

$\hat{\mathbf{s}}_T \leftarrow \hat{\mathbf{s}}_T = \mathbf{s}_T + \sigma$;
    
\tcp{Generating}
$W \leftarrow W = W_G + \Delta W$;

\For{$t \leftarrow T, ..., 1$}  
{
    \lIfElse{$t > 1$}{$\mathbf{z} \sim \mathcal{N}(0, \mathbf{I})$}{$\mathbf{z} \leftarrow 0$}
    
    $\hat{\mathbf{s}}_{t-1} = \frac{1}{\sqrt{\alpha_t}} \big(\hat{\mathbf{s}} - \frac{1-\alpha_t}{\sqrt{1-\overline \alpha_t}} \epsilon_W (\hat{\mathbf{s}}_{t}, t, c) \big) + \delta_t \mathbf z$\;
}

\Return $\hat{\mathbf{s}}_0$;

\tcp{Extracting}
$\hat{\mathbf{w}} \leftarrow \hat{\mathbf{w}} = \mathbf{Dec}(\mathbf{AS}(\hat{\mathbf{s}}_0))$;

\tcp{Fine-tuning}
Take gradient descent step on:
$\nabla_{\Delta W + \theta_{\mathbf{Enc}} + \theta_{\mathbf{Dec}}}(\mathcal{L}_{SQ} + \mathcal{L}_{WEA})$;
}

\end{algorithm}

\section{Experimental Results}

\subsection{Experimental Settings}
\subsubsection{Datasets}
\label{sec_dataset}
In order to validate the proposed \textsc{solido}, we utilize two speech datasets: LJSpeech~\cite{Ito2017ljspeech} and LibriTTS~\cite{zen2019libritts}. 
Concretely, LJSpeech is a single-speaker dataset consisting of 13,000 audio clips with a sampling rate of 22.05 kHz, while LibriTTS is a multi-speaker dataset containing approximately 586 hours of speech with a sampling rate of 24 kHz.
For experimental validation, all audio samples from the LibriTTS were resampled to 22.05 kHz to conform to diffusion models' input specifications. Furthermore, all speech samples were uniformly segmented into one-second intervals.

\subsubsection{Evaluation Metrics}
\label{sec_metrics}
We evaluated the performance of our method with different objective evaluation metrics. 
Short-Time Objective Intelligibility (STOI) \cite{taal2010stoi} predicts the intelligibility of speech. 
Mean Opinion Score of Listening Quality Objective assesses speech quality based on the Perceptual Evaluation of Speech Quality (PESQ)~\cite{recommendation2001PESQ}. 
We also conducted evaluation metrics using Structural Similarity Index Measure (SSIM) \cite{wang2004ssim}, which is a metric typically used for image quality assessment
Bit-wise accuracy (ACC) is employed to evaluate the accuracy of watermark extraction.

\subsubsection{Compared Methods}
\label{sec_sota}
To comprehensively validate the effectiveness of the proposed method, we conducted comparative experiments with various watermarking techniques. 
In the experimental design, representative methods were selected from two major categories: post-hoc watermarking and generative watermarking. 
For post-hoc watermarking, three handcrafted-based methods (Normspace~\cite{saadi2019normspace}, FSVC~\cite{zhao2021desyn-fsvcm}, and PBML~\cite{natgunanathan2017patchwork}) and three deep learning-based methods (WavMark~\cite{chen2023wavmark}, AudioSeal~\cite{roman2024audioseal}, TBWM~\cite{liu2024timebre}) were chosen. 
As for generative watermarking, one typical parameter training-based method (HiFi-GANw~\cite{cheng2024hifiwm}) and one parameter frozen-based method (Groot~\cite{liu2024groot}) were selected respectively.

\begin{table}[t]
\centering
\caption{Computational Overhead}
\resizebox{0.95\linewidth}{!}{
\begin{tabular}{ccccc}
\toprule
Method       & Training Strategy & Parameters↓ & Size (MB)↓ & ACC(\%)↑ \\ 
\midrule
\multicolumn{5}{l}{\textit{Post-hoc Watermarking (Deep Learning-based)}} \\
AudioSeal~\cite{roman2024audioseal}  & FPT    & 23.33M     & 89.14   & 92.14 \\
WavMark~\cite{chen2023wavmark}       & FPT    & 2.48M      & 9.55    & 100.00 \\
TBWM~\cite{liu2024timebre}           & FPT    & 0.77M      & 33.27   & 99.98 \\
\midrule
\multicolumn{5}{l}{\textit{Generative Watermarking}} \\
HiFi-GANw~\cite{cheng2024hifiwm}     & FPT    & 13.94M     & 193.03  & 98.93 \\
Groot~\cite{liu2024groot}            & PF    & 250.28M    & 984.99 & 99.69  \\
\rowcolor[HTML]{DDDDDD} 
SOLIDO(LoRA)                         & APT    & 2.50M      & 49.88   & 98.93 \\
\rowcolor[HTML]{DDDDDD} 
SOLIDO(LoHA)                         & APT    & 2.51M      & 49.90   & 97.92 \\ 
\bottomrule
\end{tabular}
}
\label{tab_overhead}
\end{table}

\begin{table}[t]
\centering  
\caption{Ablation Study in Different Rand and Alpha of LoRA.}
\resizebox{0.95\linewidth}{!}{
\begin{tabular}{cccccccc}
\toprule
Rank & Alpha & Param.↓ & STOI↑  & PESQ↑  & SSIM↑  & ACC(\%)↑  \\
\midrule
4    & 4   & 3.84k  & 0.9425 & 2.9048 & 0.8917 & 98.17 \\
4    & 8   & 3.84k  & 0.9627 & 3.2440 & 0.8752 & 97.67 \\
\rowcolor[HTML]{DDDDDD} 
4    & 16  & 3.84k & 0.9583 & 3.2147 & 0.9056 & 98.93  \\
4    & 32  & 3.84k  & 0.9548 & 3.0058 & 0.8986 & 97.91 \\
\midrule
8    & 16  & 7.68k  & 0.9628 & 3.2274 & 0.8765 & 98.08 \\
16   & 16  & 15.36k & 0.9613 & 3.2091 & 0.8763 & 98.53 \\
32   & 16  & 30.72k & 0.9635 & 3.2376 & 0.8770 & 96.67 \\
\midrule
8    & 8   & 7.68k  & 0.9286 & 2.6133 & 0.8846 & 97.29 \\ 
32   & 32  & 30.72k & 0.9642 & 3.5104 & 0.8446 & 90.48 \\ 
40   & 40  & 38.40k & 0.9444 & 2.8199 & 0.8892 & 98.59 \\
80   & 80  & 76.80k & 0.9546 & 2.9849 & 0.8983 & 98.36 \\
\bottomrule
\end{tabular}
}
\label{tab_loraconfig}
\end{table}

\begin{table*}[]
\centering
\setlength{\belowcaptionskip}{-0.01cm}
\caption{Fidelity of the Proposed Method. ↑ indicates a higher value is more desirable.}
\resizebox{0.95\textwidth}{!}{
\begin{tabular}{ccccccccccccccc}
\toprule
\multirow{2}{*}{Model} & \multirow{2}{*}{PEFT}  & \multirow{2}{*}{Dataset} & \multicolumn{4}{c}{\textit{Generated(Non-WM) $\leftrightarrow$ Natural}} & \multicolumn{4}{c}{\textit{Watermarked $\leftrightarrow$ Generated(Non-WM)}} & \multicolumn{4}{c}{\textit{Watermarked $\leftrightarrow$ Natural}} \\
\cmidrule{4-15}
&                       &          & STOI↑              & PESQ↑              & SSIM↑             & ACC(\%)↑            & STOI↑               & PESQ↑               & SSIM↑              & ACC(\%)↑             & STOI↑              & PESQ↑              & SSIM↑              & ACC(\%)↑             \\
\midrule
\multirow{4}{*}{\rotatebox{90}{DiffWave}}  & \multirow{2}{*}{LoRA} & LJSpeech & 0.9655             & 3.5120             & 0.8453            & N/A                 & 0.9583              & 3.2147              & 0.9056             & 98.93                & 0.9618             & 3.4066             & 0.8629             & 98.93                \\
&                       & LibriTTS & 0.9583             & 2.8156             & 0.8025            & N/A                 & 0.9274              & 2.7180              & 0.8800             & 98.33                & 0.9366             & 2.8153             & 0.8395             & 98.33                \\
                           \cmidrule{2-15}
& \multirow{2}{*}{LoHA} & LJSpeech & 0.9655             & 3.5120             & 0.8453            & N/A                 & 0.9594              & 3.1175              & 0.8699             & 98.78                & 0.9620             & 3.1107             & 0.9038             & 98.78                \\
&                       & LibriTTS & 0.9583             & 2.8156             & 0.8025            & N/A                 & 0.9412              & 2.6552              & 0.8389             & 99.13                & 0.9292             & 2.6198             & 0.8767             & 99.13                \\
\midrule
\multirow{4}{*}{\rotatebox{90}{PriorGrad}} & \multirow{2}{*}{LoRA} & LJSpeech & 0.9619             & 3.2154             & 0.8695            & N/A                 &  0.9653            & 3.4325             & 0.8675            &  97.92              & 0.9669          &  3.3257       &  0.9102        &  97.92      \\
&                       & LibriTTS & 0.9375             & 2.7489             & 0.8450            & N/A                 &  0.9358           & 2.8576               &  0.8342         &  98.01     &     0.9273      &  2.7745      & 0.8791         & 98.01       \\
                           \cmidrule{2-15}
& \multirow{2}{*}{LoHA} & LJSpeech & 0.9619             & 3.2154             & 0.8695            & N/A                 & 0.9577              & 3.0540              & 0.8690             & 99.28               &  0.9602            & 3.0633             & 0.9031            &  99.43   \\
&                       & LibriTTS & 0.9375             & 2.7489             & 0.8450            & N/A                 & 0.9327              & 2.6200              & 0.8431             & 99.48                &  0.9273           & 2.5899               &  0.8763         &  99.54  \\
                           \bottomrule
\end{tabular}
}
\label{tab_fidelity}
\end{table*}

\begin{table}[t]
\centering
\caption{Comparison of Fidelity on LJSpeech. ↑ indicates a higher value is more desirable. DW and PG represent DiffWave and PriorGrad.}
\resizebox{0.95\linewidth}{!}{
\begin{tabular}{ccccc}
\toprule
Method (bps)            & STOI↑           & PESQ↑           & SSIM↑           & ACC(\%)↑        \\ \midrule
\multicolumn{5}{l}{\textit{Post-hoc Watermarking (Handcrafted-based)}} \\
Normsapce~\cite{saadi2019normspace} (32)      & 0.9646          & 2.5506          & 0.8868          & \textbf{100.00} \\
FSVC~\cite{zhao2021desyn-fsvcm} (32)          & 0.9861          & 3.7866          & 0.9560          & \textbf{100.00} \\
PBML~\cite{natgunanathan2017patchwork} (100)         & 0.9984          & 3.9977          & 0.9803          & \textbf{100.00} \\
\midrule
\multicolumn{5}{l}{\textit{Post-hoc Watermarking (Deep Learning-based)}} \\
AudioSeal~\cite{roman2024audioseal} (16)      & 0.9985          & 4.5893          & \textbf{0.9811} & 92.14           \\
WavMark~\cite{chen2023wavmark} (32)           & \textbf{0.9997} & \textbf{4.4628} & 0.9690          & \textbf{100.00} \\
TBWM~\cite{liu2024timebre} (100)              & 0.9853          & 4.0353          & 0.9388          & 99.98           \\
\midrule
\multicolumn{5}{l}{\textit{Generative Watermarking}} \\
HiFi-GANw~\cite{cheng2024hifiwm} (20)               & 0.9414          & 2.5862          & 0.9447          & 98.93           \\
Groot~\cite{liu2024groot} (100)               & 0.9605          & 3.3871          & 0.9088          & 99.69           \\
\rowcolor[HTML]{DDDDDD} 
SOLIDO(DW-LoRA) & 0.9618          & 3.4066          & 0.8629          & 98.93           \\
\rowcolor[HTML]{DDDDDD} 
SOLIDO(DW-LoHA) & 0.9620          & 3.1107          & 0.9038          & 98.78           \\
\rowcolor[HTML]{DDDDDD} 
SOLIDO(PG-LoRA) & 0.9669          & 3.3257          & 0.9102          & 97.92           \\ 
\rowcolor[HTML]{DDDDDD} 
SOLIDO(PG-LoHA) & 0.9602          & 3.0633          & 0.9031          & 99.43           \\ 
\bottomrule
\end{tabular}
}
\label{tab_comparison_fidelity}
\end{table}


\subsection{Implementation Details}
\subsubsection{Model Settings}
We employ two representative speech diffusion models, DiffWave~\cite{kong2021diffwave} and PriorGrad~\cite{lee2022priorgrad}, as generative models for validation. 
Both diffusion models were implemented using their default configurations for training and inference.
To evaluate the feasibility of the proposed method, we incorporated two PEFT approaches: LoRA~\cite{hu2021lora} and LoHA~\cite{hyeon2021loha}.
The PEFT hyperparameters were configured as follows: both LoRA and LoHA were implemented with rank $r=4$ and scaling factor $\alpha=16$. These adaptations were integrated into the final five convolutional layers of the diffusion models.
The architectural configurations were as follows: in the watermark encoder, the Conv1d layer utilized a kernel size of 3, stride of 1, and padding of 1.
For the watermark decoder, all DSC layers were implemented with a kernel size of 3, stride of 2, and padding of 1.

\subsubsection{Training Settings}
In the proposed SDFT, we employ AdamW optimizer~\cite{loshchilov2018adamw} with a learning rate of 2e-4. The batch size and epochs were set to 4 and 25, respectively.
During SDFT, we initially set $\lambda_{wea}=1$, $\lambda_{m}=0$, and $\lambda_{s}=0$. Once $\mathcal{L}_{WEA}$ dropped below a certain threshold, we updated the settings to $\lambda_{wea}=0.1$, $\lambda_{m}=0.5$, and $\lambda_{s}=0.5$.
All experiments are performed on the platform with Intel(R) Xeon Gold 5218R CPU and a NVIDIA GeForce RTX 3090 GPU.

\subsection{LoRA Parameter Configuration}
To investigate the impact of LoRA configuration on the watermark performance in the proposed method, this experiment systematically examines different settings of rank $r$ and scaling factor $\alpha$.
Specifically, we conducted three comparative experiments: (1) Evaluating watermarking performance with consistent $r$ while varying the $\alpha$. (2) Further validating the best-performing $\alpha$ from (1) across different $r$ values. (3) Rigorous examination of watermarking performance by implementing extreme parameter configurations.

Table~\ref{tab_loraconfig} delineates the comprehensive experimental outcomes.
Empirical analysis of the resultant data indicates that within Setting (1), the optimal configuration balancing speech fidelity and watermark recovery accuracy is achieved at $r=4, \alpha=16$. This equilibrium point represents a performance apex, with degradation observed when modulating $\alpha$ in either direction.
Furthermore, maintaining a constant $\alpha=16$, we investigated the impact of varying $r$ in Setting (2). 
The results revealed a clear trend: as the rank increased, the overall performance of the watermark exhibited a more pronounced degradation.
In Setting (3), even under more extreme parameter configurations, the watermark performance showed no improvement nor achieved parity with the optimal settings. Both speech quality and extraction accuracy remained significantly inferior to those in the best-case scenario.
In a nutshell, experimental results across various parameter configurations demonstrate that the optimal configure are achieved at $r=4, \alpha=16$.
Consequently, all subsequent experiments involving LoRA (or LoHA) parameters adopt this configuration.

\subsection{Computational Overhead}
\label{sec_overhead}
To elucidate the efficiency of the proposed SOLIDO in mitigating the computational overhead inherent in parameter training-based generative speech watermarking methods, we conducted a comprehensive quantitative analysis of the parametric complexity, as illustrated in Table~\ref{tab_overhead}. 
Specifically, we systematically computed and compared the total parameter count and model size across various generative watermarking approaches during the training phase.
Moreover, to provide a comprehensive comparative assessment, we extended our analysis to include deep learning-based post-hoc watermarking techniques, thereby enabling a nuanced evaluation across different watermarking paradigms.

As shown in the Table~\ref{tab_overhead}, the proposed SOLIDO requires a total of 2.50M and 2.51M trainable parameters for the LoRA and LoHA configurations, respectively. Specifically, the LoRA and LoHA introduce 3.84k and 7.68k additional parameters, while the watermark encoder and decoder maintain fixed parameter counts of 2.25M and 0.25M, consistent across both configurations.
In addition, the total model size of SOLIDO is 49.88 MB, comprising 9.55 MB for the watermark encoder-decoder, 10.11 MB for LoRA, and 30.22 MB for the diffusion model.
The proposed SOLIDO, after utilizing PEFT, requires lower computational costs and a more lightweight training process than similar generative watermarking methods, both in terms of total training parameters and model size.
Specifically, when compared to HiFi-GANw, a generative watermarking approach via FPFT, its training parameters and model size are 5.5 and 3.8 times larger than our method, respectively. 
In the case of Groot, a parameter-free training approach, despite requiring only watermark encoder and decoder training, its parameter count reaches an astonishing 250.28M—100 times that of SOLIDO. Moreover, its model size is 19 times larger than our method.
Compared to post-hoc watermarking methods, our approach requires significantly fewer training parameters than AudioSeal and is on par with WavMark. 
Although TBWM demonstrates superior parameter efficiency—attributable to its specialized focus on audio traceability functionality—it lacks the generative capabilities inherent in the SOLIDO.
Moreover, regarding model size, SOLIDO achieves a more compact overall model size (19.66 MB) when excluding the generative model, which is significantly smaller than TBWM.
In a nutshell, compared to the aforementioned methods, SOLIDO's significantly lower computational overhead further highlights its efficiency in performing generative watermarking tasks, making it practically viable for real-world watermarking tasks on diffusion models.


\begin{figure}
    \centering
    \includegraphics[width=0.95\linewidth]{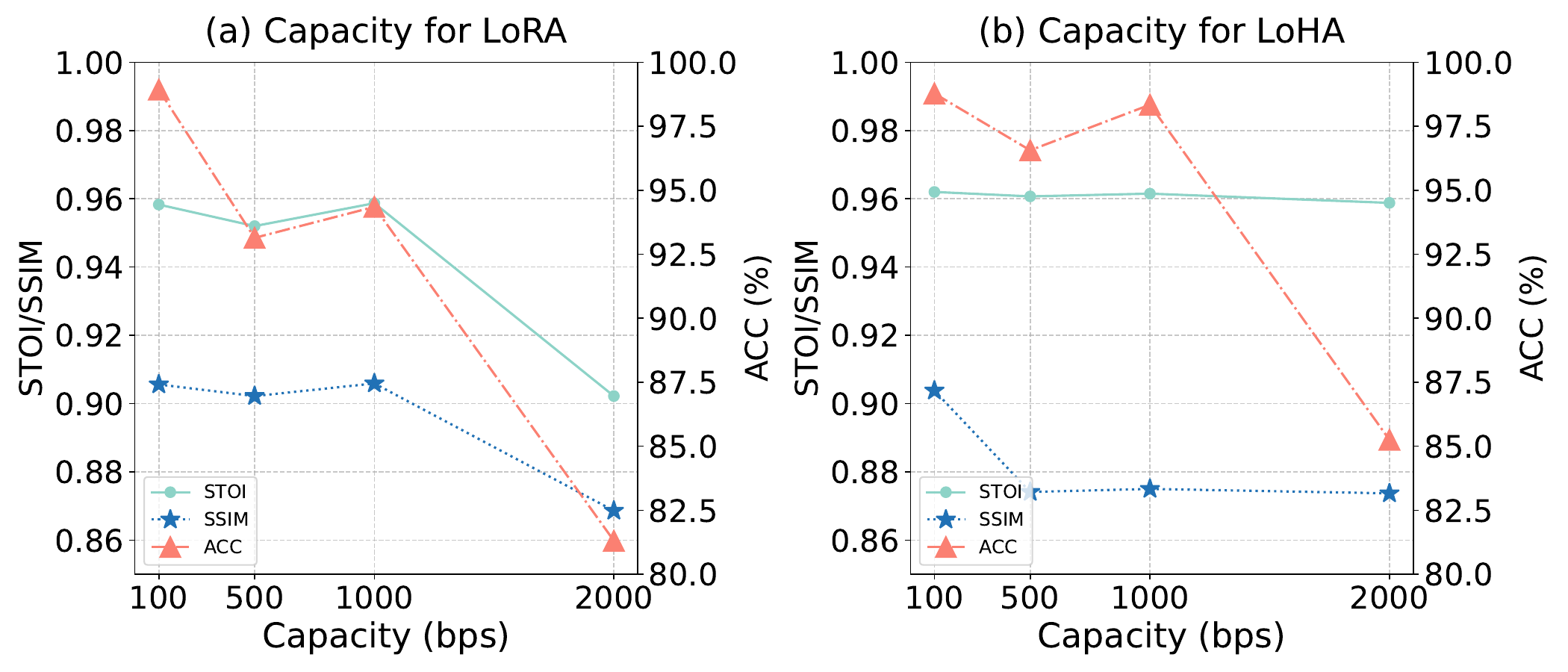}
    \caption{Capacity Analysis Across Different PEFT.}
    \label{fig_capacity}
\end{figure}


\begin{table*}[]
\centering
\caption{Robustness of the Proposed Method Against Individual Attacks. ↑ indicates a higher value is more desirable.}
\resizebox{0.95\textwidth}{!}{
\begin{tabular}{ccccccccccccccc}
\toprule
\multirow{2}{*}{Model} & \multirow{2}{*}{PEFT} & \multirow{2}{*}{Dataset} &  & \multicolumn{4}{c}{GN}            & \multicolumn{2}{c}{PN} & LPF & BPF  & Echo    & RSC & Dither  \\ 
\cmidrule{5-15} 
&   &   &   & 5 dB   & 10 dB  & 15 dB  & 20 dB  & 0.3        & 0.5       & 3 kHz   & 0.3-8 kHz & Default & 50\%        & Default \\ 
\midrule
\multirow{12}{*}{\rotatebox{90}{DiffWave~\cite{kong2021diffwave}}}  & \multirow{6}{*}{LoRA} & \multirow{3}{*}{LJSpeech} & PESQ↑    & 1.0372 & 1.0841 & 1.2235 & 1.5352 & 1.1943     & 1.0866    & 4.6185  & 3.7717    & 1.1702  & N/A         & 4.6439  \\
&                       &                           & SSIM↑    & 0.4589 & 0.5376 & 0.6202 & 0.7030 & 0.6633     & 0.5784    & 0.8846  & 0.7891    & 0.5706  & N/A          & 1.0000  \\
&                       &                           & ACC(\%)↑ & 98.34  & 98.66  & 98.85  & 98.90  & 98.55      & 98.18     & 98.51   & 98.89     & 97.25   & 96.57        & 98.81  \\
\cmidrule{3-15} 
&                       & \multirow{3}{*}{LibriTTS} & PESQ↑    & 1.0756 & 1.1820 & 1.4297 & 1.8902 & 1.1875     & 1.1022    & 4.6140  & 3.5055    & 1.2534  & N/A          & 4.6438  \\
&                       &                           & SSIM↑    & 0.4881 & 0.5672 & 0.6485 & 0.7281 & 0.6228     & 0.5642    & 0.8809  & 0.7741    & 0.5652  & N/A          & 0.9999  \\
&                       &                           & ACC(\%)↑ & 97.77  & 98.33  & 98.49   & 98.49   & 98.51      & 96.48     & 97.48   & 98.03     & 96.85   & 91.26        & 98.33  \\
\cmidrule{2-15} 
& \multirow{6}{*}{LoHA} & \multirow{3}{*}{LJSpeech} & PESQ↑    & 1.0449 & 1.1103 & 1.2961 & 1.6976 & 1.2356     & 1.1029    & 4.6180  & 3.7993    & 1.1683  & N/A          & 4.6439  \\
&                       &                           & SSIM↑    & 0.5001 & 0.5840 & 0.6701 & 0.7532 & 0.7148     & 0.6313    & 0.8923  & 0.7857    & 0.5740  & N/A          & 1.0000  \\
&                       &                           & ACC(\%)↑ & 98.89  & 98.93  & 98.90  & 99.05  & 98.56      & 97.81     & 98.69   & 98.71     & 96.26   & 96.87        & 98.73  \\ 
\cmidrule{3-15} 
&                       & \multirow{3}{*}{LibriTTS} & PESQ↑    & 1.0891 & 1.2209 & 1.5192 & 2.0476 & 1.2499     & 1.1160    & 4.6174  & 2.6387    & 2.6358  & N/A          & 4.6437  \\
&                       &                           & SSIM↑    & 0.5152 & 0.5961 & 0.6790 & 0.7582 & 0.6839     & 0.6003    & 0.8800  & 0.8440    & 0.8438  & N/A          & 1.0000  \\
&                       &                           & ACC(\%)↑ & 98.76  & 98.81  & 99.04  & 99.11  & 98.40      & 97.38     & 99.10   & 98.98     & 97.12   & 97.66        & 99.02  \\ 
\midrule
\multirow{12}{*}{\rotatebox{90}{PriorGrad~\cite{lee2022priorgrad}}} & \multirow{6}{*}{LoRA} & \multirow{3}{*}{LJSpeech} & PESQ↑    & 1.0379 & 1.0867 & 1.2274 & 1.5411 & 1.2067     & 1.0913    & 4.6186  & 3.8093    & 1.1875  & N/A          & 4.6439  \\
&                       &                           & SSIM↑    & 0.4610 & 0.5408 & 0.6243 & 0.7075 & 0.6695     & 0.5858    & 0.8860  & 0.7900    & 0.5762  & N/A          & 1.0000  \\
&                       &                           & ACC(\%)↑ & 97.30  & 97.37  & 97.88  & 97.61  & 96.89      & 96.10     & 97.81   & 97.10     & 95.72   & 95.15        & 98.02  \\ 
\cmidrule{3-15} 
&                       & \multirow{3}{*}{LibriTTS} & PESQ↑    & 1.0700 & 1.1652 & 1.3957 & 1.8327 & 1.2049     & 1.0952    & 4.6132  & 3.5013    & 1.2523  & N/A          & 4.6438  \\
&                       &                           & SSIM↑    & 0.4794 & 0.5571 & 0.6377 & 0.7171 & 0.6322     & 0.5485    & 0.8724  & 0.7747    & 0.5596  & N/A          & 0.9999  \\
&                       &                           & ACC(\%)↑ & 97.66  & 97.52  & 98.02   & 98.18  & 95.93      & 94.35     & 97.58   & 96.98     & 96.70   & 97.42        & 97.96  \\ 
\cmidrule{2-15} 
& \multirow{6}{*}{LoHA} & \multirow{3}{*}{LJSpeech} & PESQ↑    & 1.0447 & 1.1099 & 1.2946 & 1.6932 & 1.2387     & 1.1044    & 4.6175  & 3.7846    & 1.1716  & N/A          & 4.6439  \\
&                       &                           & SSIM↑    & 0.5005 & 0.5840 & 0.6699 & 0.7532 & 0.7154     & 0.6328    & 0.8919  & 0.7858    & 0.5764  & N/A          & 1.0000  \\
&                       &                           & ACC(\%)↑ & 99.27  & 99.35  & 99.18  & 99.16  & 99.44      & 99.07     & 99.22   & 99.35     & 99.51   & 98.36        & 99.28  \\
\cmidrule{3-15} 
&                       & \multirow{3}{*}{LibriTTS} & PESQ↑    & 1.0888 & 1.2182 & 1.5159 & 2.0450 & 1.2491     & 1.1157    & 4.6132  & 3.4731    & 1.2595  & N/A          & 4.6438  \\
&                       &                           & SSIM↑    & 0.5138 & 0.5947 & 0.6773 & 0.7568 & 0.6838     & 0.5985    & 0.8797  & 0.7719    & 0.5714  & N/A          & 1.0000  \\
&                       &                           & ACC(\%)↑ & 99.04  & 99.34  & 99.44  & 99.60  & 98.96      & 98.42     & 99.06   & 99.39     & 98.25   & 98.41        & 99.48  \\ 
\bottomrule
\end{tabular}
\label{tab_indi_robust}
}
\end{table*}

\begin{table*}[]
\centering
\renewcommand\arraystretch{0.9}
\caption{Comparison of ACC(\%) for Robustness Against Individual Attacks. The best results are indicated in \textbf{BOLD}, while the second-best results are \underline{underlined}.}
\resizebox{0.95\textwidth}{!}{
\begin{tabular}{ccccccccccccc}
\toprule
\multicolumn{2}{c}{\multirow{2}{*}{Method (bps)}}    & \multicolumn{4}{c}{GN}                            & \multicolumn{2}{c}{PN}           & LPF         & BPF        & Echo            & RSC        & Dither          \\ 
\cmidrule{3-13} 
\multicolumn{2}{c}{} & 5 dB            & 10 dB           & 15 dB & 20 dB & 0.3            & 0.5             & 3 kHz           & 0.3-8 kHz       & Default         & 50\%          & Default         \\ 
\midrule
\multicolumn{6}{l}{\textit{Post-hoc Watermarking (Handcrafted-based)}} \\
& \multicolumn{1}{r}{Normspace (32)~\cite{saadi2019normspace}}     & 52.64           & 58.56           & 54.17  & 62.08  & 47.31           & 47.33           & 58.06           & 52.52           & 56.30           & 49.91           & 65.75           \\
& \multicolumn{1}{r}{FSVC (32)~\cite{zhao2021desyn-fsvcm}}         & 66.49           & 73.12           & 81.08  & 88.35  & 85.10           & 81.64           & 85.86           & 75.22           & 79.76           & 52.11           & \textbf{100.00} \\
& \multicolumn{1}{r}{PBML (100)~\cite{natgunanathan2017patchwork}}  & 56.18           & 75.04           & 66.98  & 71.76  & 74.63           & 70.60           & 97.41           & 75.04           & 69.95           & 50.28           & 98.09           \\
\midrule
\multicolumn{6}{l}{\textit{Post-hoc Watermarking (Deep Learning-based)}} \\
& \multicolumn{1}{r}{AudioSeal (16)~\cite{roman2024audioseal}}    & 60.48           & 60.86           & 62.38 & 66.00  & 68.72           & 65.71& 91.64           & 70.20           & 72.77                      & 64.60           & 59.46           \\
& \multicolumn{1}{r}{WavMark (32)~\cite{chen2023wavmark}}          & 51.03           & 52.32           & 56.27  & 65.23  & 81.70           & 69.62           & \textbf{99.99}  & \textbf{99.95}  & 86.68           & -               & \textbf{100.00} \\
& \multicolumn{1}{r}{TBWM (100)~\cite{liu2024timebre}}              & 55.98           & 63.37           & 72.25  & 81.92  & 81.96           & 73.24           & \underline{99.43} & 98.83           & 94.30           & \textbf{99.31}  & \underline{99.99} \\
\midrule
\multicolumn{6}{l}{\textit{Generative Watermarking}} \\
& \multicolumn{1}{r}{HiFi-GANw (20)~\cite{cheng2024hifiwm}}     & 61.29           & 79.57           & 92.67 & 96.78  & 97.28           & 95.88  & 91.39           & 97.43           & 98.37                      & \underline{98.93}           & 98.93           \\
& \multicolumn{1}{r}{Groot (100)~\cite{liu2024groot}}               & \underline{99.13} & \textbf{99.39}  & \textbf{99.65}  & \textbf{99.37}  & \underline{99.22}           & \underline{99.04} & 98.66           & \underline{99.39} & \underline{98.67} & -               & 99.56           \\
\rowcolor[HTML]{DDDDDD} 
& \multicolumn{1}{r}{SOLIDO(DW-LoRA)} & 98.34           & \underline{98.66} & \underline{98.85} & \underline{98.90} & 98.55          & 98.18           & 98.51           & 98.89           & 97.25           & 96.57 & 98.81           \\
\rowcolor[HTML]{DDDDDD} 
& \multicolumn{1}{r}{SOLIDO(DW-LoHA)} & 98.89           & \underline{98.93} & \underline{98.90} & \underline{99.05} & 98.56          & 97.81           & 98.69           & 98.71           & 96.26           & 96.87 & 98.73           \\
\rowcolor[HTML]{DDDDDD} 
& \multicolumn{1}{r}{SOLIDO(PG-LoRA)} & 97.30           & \underline{97.37} & \underline{97.88} & \underline{97.61} & 96.89          & 96.10           & 97.81           & 97.10           & 95.72           & 95.15 & 98.02           \\
\rowcolor[HTML]{DDDDDD} 
& \multicolumn{1}{r}{SOLIDO(PG-LoHA)} & \textbf{99.27}  & \underline{99.35} & \underline{99.18} & \underline{99.16}  & \textbf{99.44} & \textbf{99.07}  & 99.22          & 99.35           & \textbf{99.51}  & 98.36 & 99.28           \\ 
\bottomrule
\end{tabular}
\label{tab_comp_indi_robust}
}
\end{table*}

\subsection{Fidelity and Capacity}
\subsubsection{Analysis of Fidelity}
\label{sec_fdl}
In this experiment, we apply the proposed \textsc{solido} to two diffusion models, DiffWave~\cite{kong2021diffwave} and PriorGrad~\cite{lee2022priorgrad}, to validate fidelity. For each model, we evaluate the quality of watermarked speech generated utilizing both LoRA~\cite{hu2021lora} and LoHA~\cite{hyeon2021loha}. Table~\ref{tab_fidelity} presents the fidelity experimental results under these settings across two datasets, LJSpeech~\cite{Ito2017ljspeech} and LibriTTS~\cite{zen2019libritts}.

The fidelity evaluation framework is systematically designed to assess three critical dimensions:
\begin{itemize}
    \item Case 1 conducts a comparative analysis between generated speech (without watermark) and natural speech (\textit{Natural $\leftrightarrow$ Generated(Non-WM)}), serving as a benchmark to evaluate the intrinsic generation capability of the diffusion model itself.

    \item Case 2 performs a comparative evaluation between watermarked speech and generated speech (\textit{Watermarked $\leftrightarrow$ Generated(Non-WM)}), providing a quantitative assessment of the speech quality degradation attributable solely to the watermarking process.

    \item Case 3 establishes an evaluation by comparing watermarked speech with natural speech (\textit{Watermarked $\leftrightarrow$ Natural}), comprehensively assessing the composite effects of both the generation and watermarking process on speech quality.
\end{itemize}
The rationale for this experimental design stems from the dual constraints that generative watermarking imposes on fidelity, where both the generation and watermarking simultaneously govern output quality. This fundamentally differs from post-hoc watermarking approaches, where speech quality metrics can directly and exclusively reflect watermarking performance.

The experimental results presented in the Table~\ref{tab_fidelity} demonstrate that the proposed SOLIDO successfully synthesizes watermarked speech while maximally preserving the diffusion models' inherent generation capability.
In terms of speech quality metrics, the results of Case 2 show only a minor degradation compared to Case 1, indicating that the watermarking process has a negligible impact on speech quality. Furthermore, comparing Case 3 with Case 2 reveals that after accounting for the additional influence of the generation process, the performance either remains comparable or shows only a marginal decline.
Synthesizing all experimental evidence, the proposed SOLIDO ensures that the watermarking process introduces only minimal degradation to speech generation, with the final watermarked speech maintaining objectively verified fidelity.

\subsubsection{Comparison of Fidelity}
To further comprehensively validate the fidelity of the proposed method, we conducted comparative experiments with existing watermarking methods.
Table~\ref{tab_comparison_fidelity} presents the fidelity evaluation results of different approaches at their default capacity settings.
Our SOLIDO demonstrates fidelity performance under various configurations employing different diffusion models and PEFT techniques, with all four experimental setups maintaining a consistent watermark capacity of 100 bps.
The experimental results demonstrate that the proposed SOLIDO achieves superior speech generation capability among generative watermarking methods.
While numerical comparisons in PESQ with deep learning-based post-hoc watermarking approaches still show some performance gaps, this can be explained by the additional generation process inherent to generative watermarking, as analyzed in Section~\ref{sec_fdl}.
However, when examining the other two evaluation metrics, our method demonstrates comparable performance to post-hoc approaches. A comprehensive analysis incorporating all three metrics confirms that the watermarked speech maintains considerable quality overall.

\begin{table*}[]
\centering
\renewcommand\arraystretch{0.6}
\caption{Robustness of The Proposed Method Against Compound Attacks.}
\footnotesize
\resizebox{0.95\textwidth}{!}{
\begin{tabular}{ccccccccccc}
\toprule
Model & PEFT & Dataset &          & GN+BPF  & GN+Echo & GN+Dither & GN+PN  & PN+BPF  & PN+Echo & PN+Dither \\ \midrule
\multirow{12}{*}{\rotatebox{90}{DiffWave~\cite{kong2021diffwave}}}  & \multirow{6}{*}{LoRA} & \multirow{3}{*}{LJSpeech} & PESQ↑    & 1.4787 & 1.1105  & 1.5353    & 1.0721 & 1.2561 & 1.0465  & 1.0851    \\
&                       &                           & SSIM↑    & 0.5494 & 0.4527  & 0.7029    & 0.5393 & 0.4731 & 0.3819  & 0.5779    \\
&                       &                           & ACC(\%)↑ & 99.21 & 97.44 & 99.19 & 98.51 & 98.27 & 96.15 & 98.14 \\ 
\cmidrule{3-11}
&                       & \multirow{3}{*}{LibriTTS} & PESQ↑    & 1.7684 & 1.1612  & 1.8902    & 1.0889 & 1.2712 & 1.0547  & 1.1016    \\
&                       &                           & SSIM↑    & 0.5570 & 0.4588  & 0.7281    & 0.5386 & 0.4433 & 0.3678  & 0.5643    \\
&                       &                           & ACC(\%)↑ & 98.14 & 97.22 & 98.49 & 96.43 & 96.24 & 94.11 & 95.90  \\ 
\cmidrule{2-11} 
& \multirow{6}{*}{LoHA} & \multirow{3}{*}{LJSpeech} & PESQ↑    & 1.1199 & 1.0556  & 1.1102    & 1.0451 & 1.3098 & 1.0502  & 1.1022    \\
&                       &                           & SSIM↑    & 0.4436 & 0.3884  & 0.5840    & 0.5070 & 0.5086 & 0.4148  & 0.6321    \\
&                       &                           & ACC(\%)↑ & 99.07 & 96.75 & 98.69 & 98.32 & 98.25 & 95.71 & 98.19 \\ 
\cmidrule(l){3-11} 
&                       & \multirow{3}{*}{LibriTTS} & PESQ↑    & 1.2528 & 1.0871  & 1.2209    & 1.0602 & 1.3030 & 1.0570  & 1.1183    \\
&                       &                           & SSIM↑    & 0.4409 & 0.3898  & 0.5965    & 0.5060 & 0.4674 & 0.3889  & 0.6013    \\
&                       &                           & ACC(\%)↑ & 98.92 & 96.50 & 98.87 & 97.26 & 97.12 & 95.00 & 97.53 \\ 
\midrule
\multirow{12}{*}{\rotatebox{90}{PriorGrad~\cite{lee2022priorgrad}}} & \multirow{6}{*}{LoRA} & \multirow{3}{*}{LJSpeech} & PESQ↑    & 1.0958 & 1.0492  & 1.0866    & 1.0384 & 1.2672 & 1.0484  & 1.0916    \\
&                       &                           & SSIM↑    & 0.4140 & 0.3590  & 0.5408    & 0.4636 & 0.4789 & 0.3879  & 0.5856    \\
&                       &                           & ACC(\%)↑ & 97.37  & 95.13  & 97.79  & 95.88  & 95.93   & 92.98  & 95.64  \\ 
\cmidrule(l){3-11} 
&                       & \multirow{3}{*}{LibriTTS} & PESQ↑    & 1.1966 & 1.0741  & 1.1660    & 1.0494 & 1.2531 & 1.0530  & 1.0984    \\
&                       &                           & SSIM↑    & 0.4127 & 0.3629  & 0.5571    & 0.4628 & 0.4317 & 0.3576  & 0.5516    \\
&                       &                           & ACC(\%)↑ & 97.23  & 96.12  & 97.95  & 93.97  & 94.10  & 92.85  & 93.89   \\ 
\cmidrule(l){2-11} 
& \multirow{6}{*}{LoHA} & \multirow{3}{*}{LJSpeech} & PESQ↑    & 1.1184 & 1.0549  & 1.1098    & 1.0444 & 1.3077 & 1.0507  & 1.1049    \\
&                       &                           & SSIM↑    & 0.4437 & 0.3870  & 0.5840    & 0.5056 & 0.5080 & 0.4143  & 0.6305    \\
&                       &                           & ACC(\%)↑ & 99.20 & 97.61 & 99.00 & 98.50 & 98.46 & 97.49 & 98.78 \\ 
\cmidrule(l){3-11} 
&                       & \multirow{3}{*}{LibriTTS} & PESQ↑    & 1.2481 & 1.0844  & 1.2194    & 1.0595 & 1.3048 & 1.0582  & 1.1168    \\
&                       &                           & SSIM↑    & 0.4394 & 0.3844  & 0.5949    & 0.5047 & 0.4666 & 0.3910  & 0.6003    \\
&                       &                           & ACC(\%)↑ & 99.06 & 98.12 & 99.37 & 98.43 & 98.32 & 96.97 & 98.46 \\ 
\bottomrule
\end{tabular}
}
\label{tab_compound_attack}
\end{table*}

\begin{table*}[]
\centering
\renewcommand\arraystretch{0.6}
\caption{Comparison of ACC(\%) for Robustness Against Compound Attacks. The best results are indicated in \textbf{BOLD}, while the second-best results are \underline{underlined}.}
\tiny
\resizebox{0.95\textwidth}{!}{
\begin{tabular}{ccccccccc}
\toprule
\multicolumn{2}{c}{Method (bps)}                            & GN+BPF & GN+Echo & GN+Dither & GN+PN & PN+BPF & PN+Echo & PN+Dither \\ 
\midrule
\multicolumn{8}{l}{\textit{Post-hoc Watermarking (Handcrafted-based)}} \\[2pt]
& \multicolumn{1}{r}{Normspace (32)~\cite{saadi2019normspace}}    & 50.76 & 55.06  & 59.06    & 47.03 & 50.87 & 48.30  & 47.49    \\
& \multicolumn{1}{r}{FSVC (32)~\cite{zhao2021desyn-fsvcm}}        & 72.55 & 67.27  & 73.41    & 72.37 & 79.45 & 71.20  & 79.53    \\
& \multicolumn{1}{r}{PBML (100)~\cite{natgunanathan2017patchwork}} & 60.86 & 58.89  & 60.89    & 60.13 & 70.98 & 65.66  & 70.78    \\
\midrule
\multicolumn{8}{l}{\textit{Post-hoc Watermarking (Deep Learning-based)}} \\[2pt]
& \multicolumn{1}{r}{AudioSeal (16)~\cite{roman2024audioseal}}    & 60.60 & 58.86  & 60.87    & 62.30 & 64.80 & 62.80  & 65.94    \\
& \multicolumn{1}{r}{WavMark (32)~\cite{chen2023wavmark}}         & 51.25 & 48.96  & 52.89    & 52.55 & 65.82 & 56.36  & 69.19    \\
& \multicolumn{1}{r}{TBWM (100)~\cite{liu2024timebre}}             & 59.61 & 59.35  & 59.87    & 59.07 & 70.49 & 76.75  & 80.92    \\
\midrule
\multicolumn{8}{l}{\textit{Generative Watermarking}} \\[2pt]
& \multicolumn{1}{r}{HiFi-GANw (20)~\cite{cheng2024hifiwm}}         & 78.94 & 81.15  & 79.95    & 78.79 & 93.40 & 96.02  & 95.66    \\
& \multicolumn{1}{r}{Groot (100)~\cite{liu2024groot}}              & \textbf{99.22} & \textbf{97.61}  &\textbf{99.52}    & \textbf{98.61} & \textbf{99.41} & \textbf{97.66} & \textbf{99.26} \\
\rowcolor[HTML]{DDDDDD} 
& \multicolumn{1}{r}{SOLIDO(DW-LoRA)}                        & \underline{99.21} & \underline{97.44}   & \underline{99.19}     & \underline{98.51} & \underline{98.27} & \underline{96.15}   & \underline{98.14}     \\
\rowcolor[HTML]{DDDDDD} 
& \multicolumn{1}{r}{SOLIDO(DW-LoHA)}                        & \underline{99.07} & \underline{96.75}   & \underline{98.69}     & \underline{98.32} & \underline{98.25} & 95.71   & \underline{98.19}     \\
\rowcolor[HTML]{DDDDDD} 
& \multicolumn{1}{r}{SOLIDO(PG-LoRA) }                       & \underline{97.37} & \underline{95.13}   & \underline{97.79}     & \underline{95.88} & \underline{95.93} & 92.98   & 95.64     \\
\rowcolor[HTML]{DDDDDD} 
& \multicolumn{1}{r}{SOLIDO(PG-LoHA)}                        & \underline{99.20} & \textbf{97.61}   & \underline{99.00}     & \underline{98.50} & \underline{98.46} & \underline{97.49}   & \underline{98.78}     \\ 
\bottomrule
\end{tabular}
}
\label{tab_comp_compound_attack}
\end{table*}

\subsubsection{Analysis of Capacity}
We evaluated the capacity performance of our \textsc{solido} by validating it on DiffWave using LoRA and LoHA at capacities of 100, 500, 1000, and 2000 bps, respectively. All experiments were conducted on the LJSpeech dataset.
Figure 3 illustrates the relationship between the capacity and performance metrics of our proposed method.
For the LoRA implementation, we observe a notable decline in accuracy at 500 bps, despite maintaining comparable speech quality. The performance metrics at 1000 bps remain largely consistent with those at 500 bps. However, increasing the capacity to 2000 bps results in substantial deterioration of both speech quality and extraction accuracy relative to 1000 bps, though the accuracy remains robust at 87\%.
In contrast, the LoHA implementation exhibits remarkable stability at lower capacities, with performance at 500 bps and 1000 bps remaining virtually identical to that at 100 bps, with only SSIM showing a minor degradation. At the elevated capacity of 2000 bps, while speech quality metrics remain relatively stable, extraction accuracy experiences a considerable decrease, yet maintains effectiveness above 85\%.
The evidence presented above demonstrates that our method achieves a maximum operational capacity of 2000 bps, beyond which performance metrics indicate significant degradation in both speech quality and watermark extraction reliability.

\subsection{Robustness}

\subsubsection{Robustness Against Individual Attacks}
To validate the robustness against individual attacks, we conducted evaluations using the seven types of attacks listed in Table~\ref{tab_attack}.
Moreover, the attack intensities configured during validation were set significantly higher than those in the attack simulator.
Table~\ref{tab_indi_robust} presents the robustness evaluation of the proposed method applied to DiffWave and PriorGrad, using both LoRA and LoHA configurations (denoted as D-R, D-H, P-R, and P-H, respectively). The results were validated across the LJSpeech and LibriTTS datasets.

Experimental results demonstrate that the proposed method maintains excellent robustness against multiple individual attacks across all four schemes (D-R, D-H, P-R, and P-H).
Specifically, under the D-R scheme, the robustness patterns remained consistent across LJSpeech and LibriTTS datasets, achieving average ACC values of 99.17\% and 97.27\% against all individual attacks.
Despite replacing LoRA with LoHA for the D-H scheme, the proposed method maintained strong robustness, attaining average ACC values of 98.31\% and 98.49\% on two datasets, respectively.
In addition, under the P-R scheme, the proposed method achieved average ACC values of 97.00\% and 97.12\% on two datasets, respectively. 
Moreover, the P-H scheme demonstrated superior robustness, attaining higher average ACC values of 99.20\% and 99.04\% on the respective datasets.
The analysis across the aforementioned four schemes conclusively demonstrates that the proposed SOLIDO effectively integrates PEFT (LoRA and LoHA) with watermarking technology, while maintaining strong resistance against individual attacks across different diffusion models.

Furthermore, to comprehensively highlight the robustness of the proposed method, we conducted comparative experiments with existing SOTA approaches, with detailed results presented in Table~\ref{tab_comp_indi_robust}.
The experimental data conclusively demonstrates that our method achieves excellent robustness compared to SOTA approaches.
When confronting Gaussian noise, while our method slightly underperforms Groot under high-decibel noise attacks, SOLIDO demonstrates the most robust performance at the maximum noise intensity (5 dB). Simultaneously, SOLIDO also achieves the highest ACC when dealing with pink noise.
When addressing other individual attacks, although our SOLIDO does not guarantee the highest ACC among all methods, it still achieves exceptionally high ACC, and its performance is only slightly lower than that of the best-performing method.
In summary, the proposed SOLIDO not only demonstrates consistent robustness across all four configurations but also ranks among the top-performing approaches when compared to SOTA methods.


\begin{table}[t]
\centering
\caption{Impact of Variable-Length Attacks for Different Watermarking Techniques.}
\resizebox{0.95\linewidth}{!}{
\begin{tabular}{crccc}
\toprule
\multicolumn{2}{c}{Method} & Variable-length & RSC & TS\\ 
\midrule
& WavMark~\cite{chen2023wavmark}      & \ding{56}      & -      & -       \\
& DeAR~\cite{liu2023dear}             & \ding{56}      & -      &  -      \\
& Groot~\cite{liu2024groot}           & \ding{56}      & -      & -       \\
& AudioSeal~\cite{roman2024audioseal} & \ding{52}      & 64.60  & 58.46   \\
& TBWM~\cite{liu2024timebre}          & \ding{52}      & 99.31  &  49.52  \\
& HiFi-GANw~\cite{cheng2024hifiwm}    & \ding{52}      & 98.93  & 52.51   \\
\rowcolor[HTML]{DDDDDD} 
& SOLIDO (Ours)                       & \ding{52}      & 96.57  & 81.81   \\
\bottomrule
\end{tabular}
}
\label{tab_variable}
\end{table}

\subsubsection{Robustness Against Compound Attacks}
In real-world scenarios, practical environments often involve more complex and severe unknown attacks. To meet the requirements of actual application conditions, we further conducted robustness experiments on the proposed method against compound attacks. 
Specifically, compound attacks were formed by arbitrarily combining two individual attacks. 
For this experiment, as illustrated in Table~\ref{tab_compound_attack}, we configured seven specific composite attacks as follows: (1) GN+BPF, (2) GN+Echo, (3) GN+Dither, (4) GN+PN, (5) PN+BPF, (6) PN+Echo, (7) PN+Dither.
The parameter configurations for each compound attack remain consistent with those used in individual attacks.

The experimental results demonstrate that the proposed SOLIDO still maintains strong robustness against composite attacks.
Specifically, under the D-R scheme, it achieves average ACCs of 98.13\% and 96.65\% on LJSpeech and LibriSpeech, respectively. Similarly, the D-H scheme demonstrates comparable performance with average ACC values of 97.85\% and 97.31\% on the respective datasets.
For PriorGrad, the P-R scheme achieves average ACC rates of 95.82\% and 95.16\% on two datasets, while the P-H scheme maintains even higher performance with average ACC values of 98.43\% and 98.39\% on the respective datasets.
Our findings reveal that the P-R scheme exhibits degraded performance against both individual and compound attacks. This degradation likely stems from using LoRA parameters optimized for DiffWave, which may conflict with PriorGrad's distinct structural characteristics, ultimately leading to suboptimal adaptation and reduced effectiveness.

In addition, we conducted a comprehensive comparative analysis by evaluating SOLIDO against existing methods under identical compound attack configurations as described above, with detailed results presented in Table~\ref{tab_comp_compound_attack}.
Experimental results reveal that most SOTA methods demonstrate inadequate robustness against compound attacks, whereas our SOLIDO method exhibits excellent resilience.
Although Groot demonstrates the most exceptional performance against compound attacks, Table~\ref{tab_overhead} reveals that its superior capability comes at the cost of excessive parameter size and prohibitive training overhead. Notably, SOLIDO's ACC metrics are only marginally lower than Groot's, without exhibiting significant performance gaps.
Moreover, when compared to HiFi-GANw (another generative watermarking approach), SOLIDO demonstrates superior robustness across multiple compound attack types.
In conclusion, the proposed SOLIDO exhibits strong robustness against compound attacks, highlighting its suitability for deployment in real-world scenarios.

\subsection{Importance of Handling Variable-Length Inputs}
\label{sec_variable}
Recent advances in speech editing technology have facilitated more accessible speech manipulation (\textit{i.e.,} cropping the silent frame of speech).
It has heightened the demand for watermarking methods capable of handling variable-length inputs.
Since handcrafted-based watermarking methods do not involve variable-length input challenges, this experiment exclusively compares the remaining five deep learning-based watermarking approaches, as illustrated in Table~\ref{tab_variable}.
Specifically, the experiment was designed to include the RSC attack, followed by padding, interpolation, and another time-stretching (TS) attack.

Among these approaches, both WavMark and Groot lack the capability to process variable-length inputs, rendering them unsuitable for direct application to trimmed speech segments.
In contrast, both TBWM and our \textsc{solido} successfully extract watermarks from variable-length speech inputs.
We further investigated the watermark extraction capability under two distinct operations: zero-padding and interpolation.
Moreover, when confronting TS, only our method is still capable of effectively extracting the watermark, while the other methods, regardless of whether they can handle variable-length inputs, lack sufficient robustness to cope with such distortions.
The experimental results demonstrate that while padding and interpolation can partially mitigate the limitations of fixed-length inputs, the extraction accuracy achieved through these operations significantly declines, particularly with the most commonly used interpolation.
In conclusion, the capability to accommodate variable-length inputs enables watermarking methods to process speech segments of diverse durations with flexibility, thereby facilitating more real-world applications.

\begin{figure}
    \centering
    \includegraphics[width=0.85\linewidth]{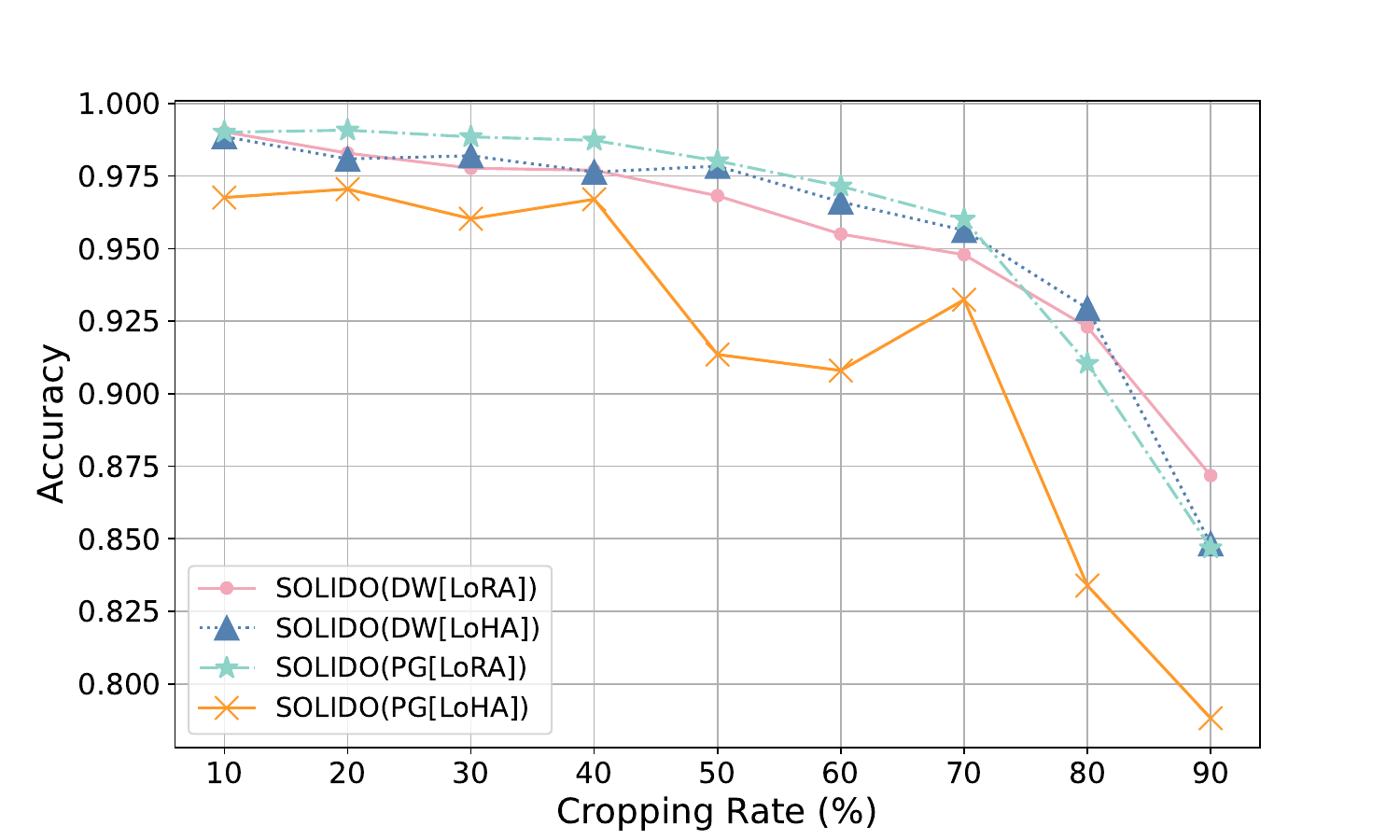}
    \caption{Robustness against various rates of rear-segment cropping attacks.}
    \label{fig_cropping}
\end{figure}

\section{Conclusion}
In the era of AIGC, there is an urgent necessity to develop sustainable and easily deployable watermarking solutions.
To this end, we first investigate the intersections of PEFT and speech watermarking and propose a generative speech watermarking method via LoRA.
Specifically, a watermark encoder is designed to transform the watermark into the latent variables aligned with the input of the diffusion model.
For high-precision watermark extraction, a watermark decoder based on DSCs is utilized to capture fine-grained temporal features from the speech waveform.
Furthermore, we propose a speech-driven lightweight fine-tuning strategy that introduces additional parameters to the generative model while keeping its original parameters frozen, achieving high-fidelity speech generation with significantly reduced computational overhead.
Extensive experiments demonstrate our SOLIDO's high fidelity across varying capacity settings, while robustness validation confirms its exceptional performance against both individual and composite attacks—particularly its resilience to variable-length attacks.

\bibliographystyle{IEEEtran}
\bibliography{ref.bib}

\end{document}